\documentstyle[12pt]{article}
\newcommand{\be}{\begin{equation}}
\newcommand{\ee}{\end{equation}}

\newcommand{\bea}{\begin{eqnarray}}
\newcommand{\eea}{\end{eqnarray}}
\newcommand{\ba}{\begin{array}}
\newcommand{\ea}{\end{array}}
\newcommand{\lsim}{{\;\raise0.3ex\hbox{$<$\kern-0.75em\raise-1.1ex\hbox{$\sim$}}\;}}
\newcommand{\gsim}{{\;\raise0.3ex\hbox{$>$\kern-0.75em\raise-1.1ex\hbox{$\sim$}}\;}}

\newcommand{\newsection}[1]{
\vspace{10mm}
\pagebreak[3]
\addtocounter{section}{1}
\setcounter{footnote}{0}
\begin{flushleft}
{ \bf{ \thesection. #1}}
\end{flushleft}
\nopagebreak
\vspace{4mm}
\nopagebreak}

\newcommand{\NP}[1]{Nucl. Phys.\ {\bf #1}\ }
\newcommand{\PL}[1]{Phys. Lett.\ {\bf #1}\ }

\newcommand{\PR}[1]{Phys. Rev.\ {\bf #1}\ }

\newcommand{\CR}{\nonumber \\}

\newcommand{\cL}{{\cal L}}

\newcommand{\half}{{1\over2}}
\newcommand{\del}{\partial}

\newcommand{\D}{\delta}
\newcommand{\DE}{\Delta}

\newcommand{\T}{\theta}

\newcommand{\ssu}{$SU(2)_L\times SU(2)_R\times U(1)_{B-L}\,$}
\newcommand{\sul}{$SU(2)_L$}
\newcommand{\sulu}{$SU(2)_L\times U(1)_Y$}
\newcommand{\sur}{$SU(2)_R$}
\newcommand{\matr}{\left( \begin{array}}
\newcommand{\ematr}{\end{array} \right)}

\newcommand{\dis}{\displaystyle}

\newcommand{\lr}{{left-right symmetric model}}

\begin{document}

\begin{titlepage}

\mbox{}\vspace*{-1cm}\hspace*{9cm}\makebox[7cm][r]{\large  HU-SEFT R 1993-16}
\vfill

\Large

\begin{center}
{\bf Supersymmetric left-right  model and its  tests in linear colliders}

\bigskip
\normalsize
{${\rm K. Huitu,}^a\:\: {\rm J. Maalampi}^b\:\: {\rm and} \:\:  {\rm M.
Raidal}^{a,b}$\\[15pt] $^a${\it Research Institute for High Energy Physics,
University of Helsinki}\\$^b${\it Department of Theoretical Physics,
University of Helsinki}}


\bigskip

\today

\vfill

\normalsize

{\bf\normalsize \bf Abstract}

 \end{center}

\normalsize

We investigate phenomenological implications of a supersymmetric left-right
model
based on \ssu gauge symmetry testable in the next generation linear colliders.
We concentrate in particular on the doubly charged $SU(2)_R$ triplet higgsino
$\tilde\Delta$, which
we find very suitable for experimental search.  We estimate its production rate
    in $e^+e^-$,
$e^-e^-$, $e^-\gamma$ and $\gamma\gamma$ collisions and consider its subsequent
decays.
These  processes have a clear discovery signature with a very low background
from other
processes.

\vfill

E-mail addresses: huitu@phcu.helsinki.fi, maalampi@phcu.helsinki.fi,
raidal@phcu.helsinki.fi

\end{titlepage}

\newpage

\setcounter{page}{2}

\newsection{Introduction}
Among the possible extensions of the Standard Model of electroweak interactions
perhaps  the most appealing one is the \lr\  based on
the gauge group \ssu\ \cite{pati}.
Apart from its original motivation of  providing a dynamical explanation for
the parity violation observed in  low-energy weak interactions,
this model differs
from the Standard  Model in another important respect:
it can explain the observed
lightness of neutrinos in a natural way. Neutrino masses are created through
the
see-saw mechanism \cite{seesaw}, according to which there are in each family a
light neutrino, much lighter than the charged fermions of the family,
and a heavy neutrino.
The anomalies measured in the
solar \cite{sun} and atmospheric \cite{atmos} neutrino fluxes seem indeed
to indicate that neutrinos should have a small but non-vanishing mass.
Furthermore,
the recent observations of the COBE satellite \cite{cobe} may indicate that
there
exists a hot neutrino component in the dark matter of the Universe. The see-saw
mechanism can account for all these phenomena, while in the Standard Model
neutrinos are  massless. In other respects the \lr\  in the low-energy limit is
very similar to the  Standard Model  and is like it  in a good agreement with
all  the laboratory experiments  performed so far.

On the technical side, the \lr\  has a naturality problem similar to that
in the Standard Model: the masses of the fundamental  Higgs scalars diverge
quadratically. To make these divergences  cancel one has to fine tune
the parameters of the theory to some 28 decimal places. As in the Standard
Model, the
supersymmetry (susy) can be used to stabilize the scalar masses and cure  this
hierarchy problem. There are also other arguments in favor of supersymmetry. It
may, for example, play  a fundamental role in the theory of quantum gravity.

In this paper we shall study some phenomenological aspects of a supersymmetric
extension of the \lr \footnote{Supersymmetric
left-right model has been studied also in \cite{beffV,bsusylr,saif,bKM}.}.   So
far
there are no  experimental evidence for  the right-handed interactions
predicted by the \ssu\ theory, let alone supersymmetry.
Nevertheless, these concepts have so many attractive features that they
deserve an experimental
and  phenomenological scrutiny. The next generation linear electron colliders
\cite{wiik} will provide an excellent environment for such a study
 as they are planned to operate in the
 energy range
from 0.5 to 2 TeV where new phenomena, such as  left-right symmetry and
supersymmetry,  are expected to manifest
themselves.

 The left-right symmetric model
itself, without supersymmetry, has many interesting predictions,
which can be studied
in high-energy electron-positron and electron-electron collisions. These have
been recently investigated  in refs.
\cite{maalampi},\cite{others},\cite{minkowski}. In the
present paper we will concentrate on the processes, where supersymmetry is
involved. We will look for reactions distinctive for the supersymmetric
left-right model allowing to distinguish it from the non-susy theory and e.g.
the
susy version of the Standard Model. (The experimental signatures of the minimal
susy Standard Model in linear colliders have been investigated in ref.
\cite{Ruckl}.) In particular we will study the production of the susy partner
of
the doubly charged Higgs boson, a novel prediction of the model, and the
subsequent decays.

The organization of the paper is as follows. In Section 2 we define our susy
\ssu\ model. We will consider  a minimal version of
the theory, where the number of Higgs fields is the smallest possible.
It turns out
that minimal set of scalars consists of two bidoublets transforming as ({\bf
2,2},0) under \ssu, and two right-handed triplets ({\bf 1,3},2) and ({\bf
1,3},-2). In Section 3 we investigate the decays of the doubly charged triplet
higgsino and the charged sleptons to find  experimental signals of the
doubly charged triplet higgsino production. In Section 4 we consider various
processes in  linear colliders where the triplet higgsinos could be produced
and
calculate their cross sections. A discussion and conclusions are given in
Section  5.

\newsection{A Supersymmetric Left-Right Model}

Apart from the existence of the superpartners of the ordinary left-right  model
particles, the most significant difference between the ordinary and  the
supersymmetric left-right model  concerns the Higgs sector. In the non-susy
theory the minimal set of Higgs fields consists of a bidoublet
\begin{equation}
\begin{array}{c}
{\dis\phi =\matr{cc}\phi_1^0&\phi_1^+\\\phi_2^-&\phi_2^0
\ematr = ({\bf 2},{\bf 2},0),}
\end{array}
\end{equation}
and a SU(2)$_R$ triplet
\begin{equation} \begin{array}{c}
{\dis\Delta=\matr{cc}\frac{1}{\sqrt{2}}\Delta^+&\Delta^{++}\\
\Delta^0&-\frac{1}{\sqrt{2}}\Delta^+\ematr = ({\bf 1},{\bf 3},2)}.
\end{array}\end{equation}
The bidoublet breaks the \sulu\ symmetry and thereby
gives  masses to quarks and charged leptons, as well  as to light weak bosons
$W_1$ and $Z_1$. The
$W_1$ and $Z_1$ are,  up to a possible small mixing with the
right-handed counterparts, the ordinary left-handed  weak gauge bosons
associated
with the symmetry group $SU(2)_L$. The heavy and so far unobserved weak bosons
$W_2$ and $Z_2$ obtain their masses in the breaking of the $SU(2)_R\times
U(1)_{B-L}$ symmetry into $U(1)_Y$, which is caused by a non-vanishing vacuum
expectation value of the triplet Higgs field $\Delta^0$.

If one wanted to stick puritanically in the left-right symmetry of the
Lagrangian, one ought to introduce in addition to the bidoublet and the
right-handed triplet Higgs fields also a left-handed triplet Higgs multiplet
$\Delta_L=({\bf 3,1},2)$. This, however, does not have any significant role to
play in the dynamics of the theory and it can therefore be left out from the
minimal model.

How does the Higgs sector change when one moves to the supersymmetric theory?
In
supersymmetrization, the cancellation of chiral anomalies among the fermionic
partners of the triplet Higgs fields requires that  the Higgs triplet $\Delta$
is
accompanied by another triplet, $\delta$,  with opposite $U(1)_{B-L}$ quantum
number. Due to the conservation of the $B-L$ symmetry, $\delta$ does not couple
with leptons and quarks. In the model that we consider, also another bidoublet
is
added to avoid trivial  Kobayashi-Maskawa matrix for quarks. This comes about
because supersymmetry forbids a Yukawa coupling where the bidoublet appears as
conjugated. The two bidoublets will be denoted by $\phi_u$ and $\phi_d$.

We have chosen the vacuum expectation values for the Higgses, which break the
$SU(2)_L\times SU(2)_R\times U(1)_{B-L}$ into the $U(1)_{em}$,
to be as follows

\be
<\phi_u >=\left( {\begin{array}{cc} \kappa_u & 0\\ 0 & 0 \end{array}}
\right),
\: <\phi_d >=\left( {\begin{array}{cc} 0 & 0\\ 0 & \kappa_d \end{array}}
\right),\:
<\Delta>=\left( {\begin{array}{cc} 0 & 0\\ v & 0 \end{array}} \right)
 ,\:   <\delta >\equiv 0.
\label{vevs}
\ee

\noindent Here  $\kappa_{u,d}$ are of the order of the  electroweak scale
$10^2$ GeV. The vev  $v$ of the triplet Higgs has to be much larger in order to
have
the masses of the new gauge bosons $W_2$ and $Z_2$ sufficiently high. With the
choice (\ref{vevs}) of the vev's the charged gauge bosons do not mix and $W_L$
corresponds to the observed  particle. This follows from our choice of giving
to
one neutral Higgs field in both $\phi_u$ and $\phi_d$ a vev equal to zero. This
is a simplifying assumption supported by  data: the experimental upper
limit for the $W_L-W_R$ mixing angle is as small as 0.005 \cite{altarelli}.

 Whether the set (\ref{vevs}) of the vev's as such realizes the
minimization of scalar potential may actually be disputable.
This question
has been discussed in \cite{beffV,bKM}.
It was argued in \cite{beffV} that one needs to take into account
the first order radiative corrections, as well as to introduce  another pair of
Higgs triplets, in order to get at least a local minimum of the
scalar potential.
In \cite{bKM} it was noticed that for a region
in parameter space also the tree level
vacuum is stable, if also one of the remaining
electrically neutral
scalars, the superpartner of right-handed neutrino ($\tilde\nu_R$), is given a
non-zero vacuum
expectation value.
As this matter has little significance for our considerations and results,
we will in the following set for simplicity
$\langle\tilde\nu_R \rangle=0$.

Given the vev's as in Eq. (\ref{vevs}), the masses of the light weak bosons are
given by \be
m_{Z_1}  = 1/\sqrt{2} \sqrt{ (\kappa_u^2 +\kappa_d^2)(g_L^2+g^{'2})},\:\:
m_{W_1}  =   g_L \sqrt{ 1/2 (\kappa_u^2 +\kappa_d^2 )},
\ee

\vspace{0.2in}
\noindent
where $g'=g_Rg_V/\sqrt{g_R^2+g_V^2}$,
and the masses of the heavy ones by

\be
m_{Z_2}  = \sqrt{2} v \sqrt{g_R^2 + g_V^2};\:\:\:\:
m_{W_2}  =   g_R \sqrt{1/2 (\kappa_u^2 +\kappa_d^2 ) + v^2 }.
\ee

\vspace{0.2in}
\noindent
The masses of the light gauge bosons are well known from the
LEP results, $M_{W_1}=80.22\,$GeV and $M_{Z_1}=91.18\,$GeV
\cite{bPDG}. The mass constraints for the heavy weak bosons and the bounds on
the left-right mixing obtained from the low-energy charged and neutral current
data depend on the assumptions one makes.  In the case the
gauge coupling constants $g_L$ and $g_R$ of  \sul\  and
\sur, as well as the CKM-matrix and its equivalent in $V+A$ charged current
interactions, are kept unrelated, one obtains from the charged current data
the bounds \cite{phen} $g_LM_{W_2}/g_R\gsim 300$ GeV and  $g_L\zeta/g_R\lsim
0.013$, where
$\zeta$ is the $W_L-W_R$ mixing angle. From neutral current data one  can
derive the lower bound $M_{Z_2}\gsim 400$ GeV for the mass of the new $Z$-boson
and the upper bound of $0.008$ for  the $Z_1,Z_2$ mixing  angle. CDF experiment
at Tevatron  has recently obtained the mass limits $M_{W_2} > 520$ GeV and
$M_{Z_2} > 310$ GeV \cite{tevatron}.
We make the usual assumption that the left and right couplings are equal,
$g_R=g_L$.
In the numerical evaluations we take also
the vacuum expectation values $\kappa_u$ and $\kappa_d$ equal. The results we
will present are not very sensitive to these assumptions.

At the same time when the right-handed gauge symmetry is broken, the
right-handed neutrinos achieve Majorana masses via a lepton number violating
$|\Delta L|=2$ Yukawa
coupling $h_{ij}\overline{\nu^c_{iL}}\Delta^0\nu_{jR}$. The masses are    given
by
a $3\times 3$ matrix $m_M$. They may be comparable with the heavy weak boson
masses $M_{W_2}$ and $M_{Z_2}$. This leads to the see-saw mechanism which, as
mentioned, explains the smallness of the masses of the ordinary left-handed
neutrinos. The masses of the light neutrinos are given by
\be
m_{\nu}\simeq -m_Dm_M^{-1}m_D^T,
\ee
where the matrix $m_D$ follows from the Dirac-type Yukawa coupling
$f_{ij}\bar\nu_{iR}\phi\nu_{jL}$. Very little is known about the Yukawa
coupling
constants $h_{ij}$ and $f_{ij}$, but in order to have neutrino mixings they
should not be diagonal. Accordingly  the triplet Higgs and higgsino
couplings are in general flavour changing, which is an obvious advantage
concerning  the experimental discovery of these particles.

Let us now define our supersymmetric \lr.
The superpotential  is assumed to have the following form:

\bea
W & = & h_u^Q \widehat Q_L^{cT} \widehat \phi_u  \widehat Q_R
+ h_d^Q \widehat Q_L^{cT} \widehat \phi_d  \widehat Q_R \nonumber \\
&&+h_u^L \widehat L_L^{cT} \widehat \phi_u  \widehat L_R
+h_d^L \widehat L_L^{cT} \widehat \phi_d  \widehat L_R
+h_\DE \widehat L_R^{T} i\tau_2 \widehat \DE  \widehat L_R \nonumber\\
&& + \mu_1 {\rm Tr} (\tau_2 \widehat \phi_u^T \tau_2 \widehat \phi_d )
+\mu_2 {\rm Tr} (\widehat \DE \widehat \D ) .\label{pot}
\eea


\vspace{0.2in}
\noindent
Here $\widehat Q_{L(R)}$ stands for the doublet of left(right)-handed quark
superfields, $\widehat L_{L(R)}$ stands for the doublet of left(right)-handed
lepton superfields, $\widehat \phi_u$ and $\widehat \phi_d$ are the two
bidoublet
Higgs superfields, and $\widehat \DE$ and $ \widehat \D$ the two triplet
Higgs superfields. The generation indices of the quark and lepton superfields
are
not shown. The quantum numbers of the superfields  are summarized in Table 1.
In our numerical examples we will use for the Yukawa coupling constant
$h_{\Delta}=0.3$.

In the superpotential (\ref{pot}) the $R$-parity,
$R=(-1)^{3(B-L)+2S}$, is preserved.
This ensures that the susy partners with $R=-1$ are produced in pairs and
that the lightest supersymmetric particle (LSP) is stable.
The parameters $\mu_i $ in Eq. (\ref{pot}) are supersymmetric mass parameters.
They are usually close to the  scale of the soft supersymmetry breaking
parameters in order
to preserve the naturalness of the theory \cite{haber}.
In supersymmetric models, which have also a gauge singlet Higgs
field, the $\mu $-type terms are generated by giving a vacuum expectation value
for the singlet Higgs.
We assume here that the parameters $|\mu_i|$ are of the order of the weak
scale.

{}From the superpotential we can
calculate the Yukawa interaction terms for the particles.
They are given by the general formula \cite{bHK}

\be
\cL_{\rm Yukawa}=-\half [ (\del ^2 W/\del \varphi_i\del \varphi_j)\psi_i\psi_j
+
(\del ^2 W/\del \varphi_i\del \varphi_j)^* \bar\psi_i\bar\psi_j ].
\label{Yukawa}
\ee

\vspace{0.2in}
\noindent
In this formula $\varphi_k$ denote scalar fields and $\psi_k$ fermions
of the chiral superfields.
For the scalars and the fermions of the gauge superfields there are
also non-supersymmetric mass terms, the soft breaking terms
\cite{bGG}, given by

\be
{\cL_{\rm soft}}=-\half\sum_i m_i^2 |\varphi_i|^2
- \half \sum_\alpha M_\alpha\lambda_\alpha \lambda_\alpha
+B\varphi^2+ A\varphi^3 +h.c. ,
\label{soft}\ee

\vspace{0.2in}
\noindent
where the second sum corresponds to the soft breaking
terms for gauginos.
The scalar interaction terms,
$\varphi^2 $ and $\varphi^3$, are the
quadratic and cubic interaction terms, which are allowed by gauge
symmetry for scalars.
The scalar masses are found from the scalar potential

\be
V= \sum_i\left| {\frac{\del W}{\del \varphi_i }}\right|^2 + \half \sum_\alpha
\left| g_\alpha\sum_{ij} \varphi^\dagger_i T^\alpha _{ij} \varphi_j\right| ^2
+V_{soft},
\ee
where $V_{\rm soft}$ is specified by ${\cal L}_{\rm soft}$ in Eq. (\ref{soft}).

In this work we are especially interested in the doubly charged
fermions occurring in the Higgs triplet superfields.
Their mass matrix is particularly
simple, since doubly charged higgsinos do not mix with gauginos.
{}From Eq.(\ref{Yukawa})
one finds the supersymmetric mass terms for the higgsinos,

\bea
\cL_{\rm doublet \; mass} & = & -\mu_1 [-\tilde\phi^0_{2u}\tilde\phi^0_{1d}+
\tilde\phi^+_{1u}\tilde\phi^-_{2d}+\tilde\phi^-_{2u}\tilde\phi^+_{1d} -
\tilde\phi^0_{1u}\tilde\phi^0_{2d} ] +h.c.\nonumber\\
\cL_{\rm triplet\; mass} & = & -\mu_2 [\tilde\Delta ^+ \tilde\delta^- +
\tilde\Delta ^{++} \tilde\delta^{--}
+\tilde\Delta ^0 \tilde\delta^0  ] + h.c.\label{Deltamass}
\eea

\noindent
The triplet higgsinos and Higgses have lepton number two.
Consequently the final state of the higgsino decay must also
have lepton number two in the case of $R$-parity conservation.
The interaction term which includes the strength with which the doubly charged
$\tilde\Delta$ decays to lepton and slepton is found from Eq. (\ref{Yukawa}) to
be
\be
{\cL}_{\tilde\Delta\tilde ll}=-2h_{\Delta}\overline{l^c}\tilde\Delta\tilde l.
\ee
The other interactions of the doubly charged higgsinos are found from the
superfield interaction term $\widehat \varphi^\dagger e^{2g\widehat {\cal V}}
\widehat \varphi _{|\T \T\bar\T\bar\T}$
between the matter superfields $\widehat \varphi$  and
gauge superfield $\widehat {\cal V}$, and they are given by \cite{bHK}

\be
\cL_{{\rm int, gauge-matter}} =
-igT^a_{ij}V^a_{\mu}\overline\psi_i\overline\sigma^{\mu}\psi_j+
(ig_a\sqrt{2} T_{ij}^a \varphi_i^* \lambda^a \psi_j + h.c.),
\ee

\noindent
where $T$ is the generator of the gauge group.

In unbroken supersymmetry the masses of the leptons, $m_\ell $,
are equal to the masses of the sleptons.
The soft breaking terms provide new mass terms for the scalar particles
in the model.
The slepton mass matrix is of the general form \cite{bER}

\be
\left( {\begin{array}{cc} L^2\tilde m^2 + m_\ell^2  &
A\tilde m m_\ell \\  A\tilde m m_\ell  &
R^2\tilde m^2 + m_\ell^2  \end{array}}\right) ,
\ee

\vspace{0.2in}
\noindent
where $L$, $R$, and $A$ are dimensionless constants and $\tilde m$ is
a mass parameter.
These are in principle different for each generation.
When compared to the diagonal terms, the off-diagonal mixing terms
are small as they are proportional
to the lepton mass.
The experimental lower limits for the slepton masses are
approximately one half of the LEP beam energy,
$m_{\tilde \ell}> 43-45\,$GeV \cite{bPDG}.
The squark mass matrices are of the similar form. In unified supersymmetric
models the coloured states are heavier than the uncoloured sleptons
\cite{ross}.
We will assume that the squarks are much heavier than the sleptons. This
assumption will become important when one considers the decay modes of
charginos.

To find the neutralino and chargino masses we need to consider
the interaction terms between the superpartners of gauge bosons,
the Higgses, and the higgsinos.
These are given by
\bea
{\cL}_{\lambda\psi A}&=&ig_{B-L}\sqrt 2v\lambda_{B-L}\tilde\Delta^0+ig_R\sqrt 2
v(\lambda^-_R\tilde
\Delta^+-\lambda^0_R\tilde\Delta^0)+ig_L\left(\frac{\kappa_u}{\sqrt
2}\lambda_L^0
\tilde\phi^0_{1u}+\kappa_u\lambda_L^+\tilde\phi^-_{2u}\right)\nonumber\\
&&-ig_R\left(\frac{\kappa_u}{\sqrt
2}\tilde\phi^0_{1u}\lambda^0_R+\kappa_u\tilde
\phi^+_{1u}\lambda^-_R\right)+ ig_L\kappa_d\left(\lambda_L^-
\tilde\phi^+_{1d}-\frac{1}{\sqrt
2}\lambda_L^0\tilde\phi^0_{2d}\right)\nonumber\\
&&-ig_R\left( {\kappa_d}\tilde\phi^-_{2d}\lambda^+_R-\frac{\kappa_d}{\sqrt
2}\tilde \phi^0_{2d}\lambda^0_R\right) + h.c.
\eea

\noindent
The soft supersymmetry breaking terms for the gauginos can be written as

\be
\cL_{\rm soft}=-1/2 \{m_L(\lambda_L^0\lambda_L^0 +2 \lambda_L^+\lambda_L^-) +
m_R(\lambda_R^0\lambda_R^0 +2 \lambda_R^+\lambda_R^-) +
m_{B-L} \lambda_{B-L}^0\lambda_{B-L}^0 \} +h.c.
\ee

\vspace{0.2in}
\noindent
To diagonalize the chargino and neutralino mass matrices
we follow the recipe of \cite{bHK}.
We denote $\psi^{+T}=(-i\lambda^+_L,-i\lambda^+_R,
\tilde\phi_{1u}^+,\tilde\phi_{1d}^+,\tilde\Delta^+)$
and $\psi^{-T}=(-i\lambda^-_L,-i\lambda^-_R,
\tilde\phi_{2u}^-,\tilde\phi_{2d}^-,\tilde\delta^-)$.
The chargino mass matrix depends on the following parameters: the soft gaugino
masses $m_L$ and $m_R$, the supersymmetric Higgs masses $\mu_1$ and $\mu_2$,
the
vacuum expectation values $\kappa_u$, $\kappa_d$, and $v$, and the gauge
coupling $g_R$ and $g_L$. The mass Lagrangian can be written as

\be
{\cL_{\rm chargino\; mass}}=-\half (\psi^{+T} \psi^{-T}) \left(
\begin{array}{cc} 0 & X^T \\X & 0 \end{array} \right)
\left( \begin{array}{c} \psi^+\\ \psi^- \end{array} \right)
+h.c.
\ee

\vspace{0.2in}
\noindent
For a given set of values for the parameters, one can find
numerically the eigenvalues for $X^\dagger X$ and
$XX^\dagger$ matrices.
The
physical charginos $\tilde\chi^{\pm}_i,\:\: i=1,\dots 5,$ are found by
multiplying  $\psi^+$ and $\psi^-$ by the corresponding
diagonalizing matrices $C^{\pm}$:
\be
\tilde\chi^{\pm}_i=\sum_{j}C_{ij}^{\pm}\psi_j^{\pm}.
\label{charginos}
\ee
Similarly, for neutralinos we denote
$\psi^{0T}=(-i\lambda^0_L,-i\lambda^0_R,-i\lambda^0_{B-L},
\tilde\phi_{1u}^0,\tilde\phi_{2u}^0,\tilde\phi_{1d}^0,\tilde\phi_{2d}^0,
\tilde\Delta^0,\tilde\delta^0)$. The neutralino mass matrix depends in addition
to the parameters appearing in the chargino case, also on the gaugino mass
$m_{B-L}$ and the gauge coupling $g_{B-L}$. The largeness of the soft
gaugino masses determine the nature of the lightest neutralino, but are free
parameters. The measured masses of the weak vector bosons give two constraints
for the vev's and the gauge couplings. The mass Lagrangian of neutralinos  is
written as

\be
{\cL_{\rm neutralino\; mass}}=-\half \psi^{0T}Y\psi^0 +h.c.
\ee

\vspace{0.2in}
\noindent
One can then find the eigenvalues of $Y^\dagger Y$.
Multiplying the $\psi^0$ by the diagonalizing
matrix $N$ gives the
physical Majorana neutralinos $\tilde\chi^{0}_i,\:\: i=1,\dots 9$:

\be
\tilde\chi^{0}_i=\sum_jN_{ij}\psi^0_j.
\label{chargino}
\ee

For large soft gaugino masses one finds an LSP with a large higgsino component.
In the minimal supersymmetric Standard Model this is an unfavoured situation,
if one wants to solve
the dark matter problem in terms of LSP, since higgsinos annihilate too rapidly
\cite{Rosz}. In our
case, however, the large higgsino component is the triplet higgsino
$\tilde\delta^0$, for which the
cosmological situation is very different and worth a separate study.
 The chargino
and neutralino masses have also been studied in ref. \cite{saif}. In
\cite{saif}
the $\mu_2$ mixing parameter of the triplet Higgses is taken to be zero, which
would correspond to massless doubly charged higgsino.

We have calculated numerically the composition of neutralinos and charginos for
different values of the parameters.
The neutralinos are Majorana particles, whereas the charginos combine together
to form Dirac fermions.
In Table 2 we give compositions and  masses of physical charginos and
neutralinos
assuming that $m_{W_R}=500\,$GeV, the soft supersymmetry breaking parameters
are
$1\,$TeV, and $\mu_1=\mu_2=200\,$GeV.

\vfil\eject

\newsection{Decay of the triplet higgsino and slepton}

Before going to the triplet higgsino production processes we will in this
section consider its decay. The allowed decay modes are

\bea
\tilde \Delta^{++}  & \rightarrow &  \Delta^{++}\, \lambda^{0} ,\:\:
 \Delta^{+}\, \lambda^{+},\CR
&&
 \tilde \Delta^{+} \, W_2^+ ,\CR
&& \tilde l^+ l^+ .
\eea

\vspace{0.2in}
\noindent
In large regions of the parameter space, the kinematically favoured
decay mode is
$\tilde \Delta^{++}  \rightarrow  \tilde l^+ l^+ $. This is of course the
case only when $m_{\tilde l^+}<m_{\tilde\Delta^{++}}$ (at least for some lepton
flavour), which we will assume in the following.  As the   mass  of the triplet
Higgs
$\Delta$ is of the order of the SU(2)$_R$ breaking scale $v$ \cite{Gunion}, the
first two decay
channels are forbidden energetically in our case of relatively light triplet
higgsinos. For the same
reason is the channel $\tilde \Delta^{+} \, W_2^+$ kinematically disfavoured,
since the mass of
$W_2$  is known to be above 0.5 TeV. The decay channel  $\tilde \Delta^{+} \,
W_1^+$ is forbidden
in the case of no $W_L-W_R$ mixing. In the following we will assume that
$\tilde\Delta^{++}$ (and its charge conjugated state  $\tilde\Delta^{--}$)
decay
in 100\% into the $\tilde ll$ final state.

The charged leptons $\tilde l$ can decay either to a charged lepton of the same
flavour plus a neutralino, to a neutrino plus a chargino, or to a charged gauge
boson plus a
sneutrino:
\be
\tilde l^+\to l^+ + \tilde\chi_i^0,
\label{decay1}
\ee
\be
\tilde l^+\to \nu + \tilde\chi_i^+,
\label{decay2}
\ee
\be
\tilde l^+\to W^+ + \tilde\nu.
\label{decay3}
\ee
The decay mode (\ref{decay3}) is kinematically disfavoured and we do not
consider it.
As discussed earlier, there are two slepton states of a given flavour, the
left-slepton $\tilde l_L$ and the right-slepton $\tilde l_R$, which may
sligthly
mix with each other.  The decay of the mass eigenstate predominantly the
right-slepton into the neutrino channel will in general be kinematically
disfavoured or even forbidden because of the heaviness of the right-handed
neutrino.

The interaction responsible on the decays (\ref{decay1}) and (\ref{decay2}) are
given by the
Lagrangian
\bea
{\cal L}_{\tilde l{\rm -decay}}&=& \frac{1}{2\sqrt 2}\left[\bar
l(1+\gamma_5)(g_LN_{i1}+g_{B-L}N_{i3})\tilde\chi_i^0 \tilde
l_L\right.\nonumber\\
&& \left.-\bar
l(1-\gamma_5)(g_RN_{i2}^*+g_{B-L}N_{i3}^*)\tilde\chi_i^0\tilde
l_R\right]\nonumber\\
&&-\frac{1}{2}[\overline\nu(1+\gamma_5)g_LC_{i1}^{-*}\tilde\chi_i^+\tilde l_L +
\overline\nu(1-\gamma_5)g_R C_{i2}^+\tilde\chi_i^+\tilde l_R] + h.c.\nonumber\\
&&\equiv\sum_{i,j}\bar l(v_{ij}-a_{ij}\gamma_5)\chi^0_i\tilde l_j +
\sum_{i,j}\bar l(v_{ij}'-a_{ij}'\gamma_5)\tilde\chi^+_i\tilde l_j, \eea
 where $\theta$ is the mixing angle between
slepton mass eigenstates $\tilde l_1$ and $\tilde l_2$. The decay width is then
given by the formula \be
\Gamma=\frac{1}{4\pi}\sum_{i,j}(|v_{ij}|^2+|a_{ij}|^2)\frac{(m_{\tilde
l_i}^2-m_l^2-m_{\tilde\chi_j}^2)}{m_{\tilde l_i}^2}\left[\left(\frac{m_{\tilde
l_i}^2+m_l^2-m_{\tilde\chi_j}^2}{2m_{\tilde\chi_i}}\right)^2-m_l^2\right]^{1/2}. \ee

Which  of the various decay channels is the dominant one depends on the mass of
the decaying slepton. In Fig. 1  the branching ratios of the different channels
are plotted as the function of the left-slepton and right-slepton masses
(neglecting the slepton mixing). For the left-slepton decay the channel
(\ref{chargino}) becomes dominant immediately the slepton mass exceeds the
 mass of the lightest chargino. The chargino has several decay channels, e.g.
into a lepton-slepton pair, a W-chargino pair, and a quark-squark pair.

\newsection{Production of the triplet higgsino}

The next generation linear electron colliders will, besides the usual $e^+e^-$
reactions, be able to work also in $e^-e^-$, $e^-\gamma$ and $\gamma\gamma$
modes. The high energy photon beams can be obtained by back-scattering of
intensive laser beam on high energy electrons. It turns out that all these
collision modes may be useful for investigation of the susy left-right model.

In the following we shall study the following four reactions where the doubly
charged higgsinos
$\tilde\Delta^{\pm\pm}$ are produced:
\be
e^+e^-\to \tilde\Delta^{++}\tilde\Delta^{--},\label{reaction1}
\ee
\be e^-e^-\to \tilde\chi^0\tilde\Delta^{--},\label{reaction2}
\ee
\be \gamma e^-\to \tilde\l^+\tilde\Delta^{--},\label{reaction3}
\ee
\be \gamma\gamma \to \tilde\Delta^{++}\tilde\Delta^{--}.\label{reaction4}
\ee
We have chosen these reactions for investigation because they   all have a
clean
experimental signature: a few hard leptons and missing energy. Futhermore, they
all have very small background from other processes.  The fact that
$\tilde\Delta^{\pm\pm}$ carries two units of electric charge and two units of
lepton number and that it does not couple to quarks  makes the processes
(\ref{reaction1}) - (\ref{reaction4}) most suitable and distinctive tests of
the susy left-right model.

\bigskip

\noindent{\bf  Reaction $e^+e^-\to \tilde\Delta^{++}\tilde\Delta^{--}$}

\medskip

The triplet higgsino pair production in $e^+e^-$ collision occurs
 through the diagrams
presented in Fig. 2 ,  provided of course that these particles are light
enough compared with the available collision energy. In contrast with the
triplet
Higgs fields whose   mass
 is in the  TeV scale \cite{Gunion}, the
mass of the triplet higgsino, $\tilde\Delta^{\pm\pm}$,  is not strongly
constrained. What is known is that since doubly charged fermions have not been
seen in present day accelerators,  their masses cannot be much below 100 GeV.
In
the view of our theory, the mass of $\tilde\Delta^{--}$ is given by the susy
mass
parameter $\mu_2$ (see Eq. (\ref{Deltamass})), which is a free parameter. As we
mentioned before, for the reason of naturality its value should not differ too
much from the electroweak breaking scale, i.e.  $\mu_2=O(10^2\,{\rm GeV})$.

Besides the mass $M_{\tilde\Delta^{--}}$, the total cross section of the
reaction at a given collision energy depends on the unknown masses of the
selectron and the heavier neutral weak boson $Z_2$. Of course, the amplitude
of the $Z_2$ mediated reaction is strongly suppressed in comparison with
the photon exchange reaction due to the propagator effect  and thus the
$M_{Z_2}$
dependence of the cross section  is quite negligible when the experimental
lower
limit is taken into account. Note also that the reaction mediated by the
lighter
weak boson $Z_1$ is highly suppressed as  $\tilde\Delta^{--}$ couples to that
boson only through the $Z_1-Z_2$ mixing.

In Fig. 3  the total cross section for the process (\ref{reaction1}) is
presented as a function of the mass of $\tilde\Delta^{--}$ for the collision
energy of $\sqrt{s}=1$ TeV  and for two values of the selectron mass,
$m_{\tilde
l}=$ 200 GeV and  400 GeV. As can be seen, the cross section is for these
parameter values about 0.5 pb and it is quite constant up to  the threshold
region. To have an estimate for the event rate, one has to multiply the cross
section with the branching ratio of the decay channel of the produced
higgsinos
used for the search. As pointed out earlier, the favoured decay channel may be
\be
\tilde\Delta^{--}\to\tilde l^-l^-\to l^-l^-\tilde\chi^0.
\label{llchannel}\ee
Here $l$ can be any of $e,\ \mu$ and $\tau$ with practically equal
probabilities.
The importance of the competing channel with the  $\tilde\Delta^+W^+$  final
state depends on the mass of the singly charged triplet higgsino
$\tilde\Delta^+$ and the mass of $W_R$. One may assume that it is close to the
mass of the doubly charged higgsino and larger than that of the slepton $\tilde
l$, in which case the channel (\ref{llchannel}) would dominate. In any case the
signature of the pair production reaction (\ref{reaction1}) would be the purely
leptonic final state associated with  missing energy. The missing energy is
carried by neutrinos or neutralinos.

In the Standard Model a final state consisting of four charged leptons  and
missing energy can result  from cascade decays. In the susy left-right  model
there are, however, some unique final states not possible in the Standard
Model,
namely those with non-vanishing separate lepton numbers.

\bigskip

\noindent{\bf Reaction $e^-e^-\to \tilde\chi^0\tilde\Delta^{--}$}

\medskip

The production of the  triplet higgsino $\tilde\Delta^{--}$ in
electron--electron collision occurs via a selectron exchange in t-channel  (see
Fig. 4). The cross section is a function of the unknown masses
$M_{\tilde\Delta^{--}}$ and $m_{\tilde e}$. In Fig. 5 the cross section is
presented as a function of $M_{\tilde\Delta^{--}}$ for two values of the
selectron mass,  $m_{\tilde e}= $ 200 GeV and  500 GeV, at the collision energy
$\sqrt{s}= 1 $ TeV. It is taken into account in this figure that the final
state neutralino mass is
related to the triplet higgsino mass as they both depend on the parameter
$\mu_2$. The signature of
the reaction is a same-sign lepton pair created in the cascade decay
(\ref{llchannel}) of
$\tilde\Delta^{--}$, associated with the invisible energy carried by
neutralinos. As pointed out
earlier the two leptons need not be of the same flavour since the $|\Delta
L|=2$ Yukawa couplings are
not necessarily diagonal. This may be useful for distinguishing the process
from the selectron pair
production $e^-e^-\to\tilde e^-\tilde e^-\to e^-e^- +{\rm neutralinos}$, which
is the leading process
for the selectron production in the  susy version of the Standard  Model. In
the Standard
Model the final states $e^-\mu^-$, $e^-\tau^-$ and $\mu^-\tau^-$ are forbidden.

\vfil\eject

\noindent{\bf Reaction $\gamma e^-\to \tilde\l^+\tilde\Delta^{--}$}

\medskip

The mechanism for producing high-energy photon beams by Compton back-scattering
high intensity
laser pulses on high energy electron beams was proposed in ref. \cite{photon}.
The distribution of the energy fraction $y=E_{\gamma}/E_e$ transferred to the
photon
in this process is given by \cite{photon}

\be
P(y)=\frac{1}{N}(1-y+\frac{1}{1-y}-\frac{4y}{x(1-y)}+\frac{4y^2}{x^2(1-y)^2})
\ee
where
\be
x=\frac{4E_eE_{\rm laser}}{m_e^2}{\phantom{xxxxxxxx}}{\rm
and}{\phantom{xxxxxxxx}}0\leq y\leq
\frac{x}{1+x}. \ee
The factor $N$ is chosen so that $\int dyP(y)=1$. As discussed in \cite{Ruckl},
one should   tune the laser energy in such a way that $x=2(\sqrt 2+1)$, since
for higher $x$ the conversion efficiency will drop considerably due to the
possibility of the back-scattered and laser photons to produce $e^+e^-$ pairs.
As
a result, the hardest photons will have the energy about 0.83$E_e$.

There are three
Feynman diagrams contributing to the photoproduction reaction
(\ref{reaction3}):  electron exchange
in s-channel, selectron exchange in t-channel and triplet higgsino exchange in
t-channel (see Fig. 6).
In Fig. 7 the total cross section  is presented as a function of the triplet
higgsino mass for the electron-electron center of mass energy $\sqrt {s_{ee}}=$
1 TeV.   The cross
section is determined by convoluting the photon energy distribution, i.e.
$\sigma(s_{ee})=\int dyP(y)\sigma(s_{e\gamma})$.

The experimental signature of the reaction is three lepton final state
associated with missing energy. The positive lepton is any lepton, and the two
negative ones can be any combination of the electron, muon and tau, provided
the
triplet higgsino coupling is not diagonal. A suitable choice of the final state
will cut down the Standard Model background coming e.g. from the reaction
$e^-\gamma \to e^-Z*$.  The cross section is above  {\cal O}(100 fm) for a
large
range of the masses $M_{\tilde\Delta^{--}}$ and $m_{\tilde e}$, providing hence
a
good potential for the discovery of $\tilde\Delta^{--}$.

\bigskip

\noindent{\bf Reaction $\gamma\gamma \to \tilde\Delta^{++}\tilde\Delta^{--}$}

\medskip

This reaction is an alternative of, but not  competitative with, the  reaction
(\ref{reaction1}) for producing a doubly charged  higgsino pair. Feynman
diagram of the process is
presented in Fig. 8. Because the photon energies are not monochromatic but
broadly distributed, no
sharp threshold will be visible  in the production cross section. Moreover, the
maximum collision
energy will be some 20\% less than the $e^+e^-$ energy. On the other hand, the
only unknown parameter in the process is the mass $M_{\tilde\Delta^{--}}$ as
the
couplings are completely determined by the known electric charge of the
higgsino.

The cross section of the reaction as a function of $M_{\tilde\Delta^{--}}$  is
given in Fig. 9 for the collision energy $\sqrt{s_{ee}}$= 1 TeV. The
experimental
signature of the reaction will be of course the same as for the process
(\ref{reaction1}), i.e. four charged leptons associated with missing energy.
The cross section is large because of the photon coupling to electric charge.

\newsection{Discussion and conclusions}

The left-right symmetric electroweak model based on the \ssu \ symmetry has
many attractive features. In particular, in the see-saw mechanism it offers a
beautiful and very natural explanation for the lightness of the ordinary
neutrinos. On the other hand, like in the Standard Model it has a
hierarchy problem in the scalar sector, which can be solved by making the
theory supersymmetric.

We have investigated in this paper the experimental signatures of the
supersymmetric \ssu model. We have concentrated in  the production and
decay of the doubly charged $SU(2)_R$ triplet higgsino $\tilde\Delta^{++}$.
This
particle is very suitable for experimental search for many reasons. It is
doubly
charged, which means that it does not mix with other particles. Consequently
its
mass is given by a single parameter, the susy Higgs mass $\mu_2$, which has to
be positive, in contrast with $\mu_1$, which has an undetermined sign. Also the
decays of the $\tilde\Delta^{++}$ are very limited, since it carries two units
of lepton number and it does not couple to quarks.  The nonconservation of the
separate lepton
numbers $L_e$, $L_{\mu}$, and $L_{\tau}$ of the $\tilde\Delta^{++}$ couplings
may also help to
distinguish the signal from the background. These separate lepton number
violating couplings
can be studied in the slepton pair production, where one of the reaction
amplitudes includes $\tilde\Delta^{--}$ exchange \cite{letter}.

 We have calculated the production cross sections
of $\tilde\Delta^{++}$ (and $\tilde\Delta^{--}$) in $e^+e^-$, $e^-e^-$,
$e^-\gamma$ and $\gamma\gamma$ collisions. We pointed out the clear signals of
these reactions, which have no substantial background from the Standard Model
physics. From the experimental point of view the process
$\gamma\gamma\to\tilde\Delta\tilde\Delta$ is especially interesting, since it
depends only on one parameter, $\mu_2$, and its cross section is large for
$\mu_2\lsim 300$ GeV. For larger $\tilde\Delta^{++}$ masses the cross
sections are still sizable for the other processes. Depending on the situation
and
the parameters used, the cross sections are in the range 10 fb -- 1 pb.

\vspace{2cm}
\large
\noindent Acknowledgements
\normalsize
\vspace{0.5cm}

 The work has been supported by
the Finnish Academy of Science. One of the authors (M.R.) expresses his
gratitude to the
Viro-sŠŠtiš foundation for a grant.

\newpage
\noindent{\bf FIGURE CAPTIONS}

 \medskip

\noindent {\bf Figure 1.} a) Branching ratios of the left-slepton as a function
of the slepton mass
for $\mu_2=300$ GeV. b)  Branching ratios of the left-slepton as a function of
the slepton mass
for $\mu_2=120$ GeV.
c) Branching ratios of the right-slepton as a function of the slepton mass for
$\mu_2=300$ GeV.

\noindent {\bf Figure 2.} Feynman diagrams for the pair production of the
doubly charged higgsinos in
electron-positron collisions.

\noindent {\bf Figure 3.} Total cross section for the reaction $e^+e^-\to
\tilde\Delta^{++}\tilde\Delta^{--}$ as a function of the higgsino mass
$m_{\tilde\Delta^{++}}$
for two values of the selectron mass $m_{\tilde l}$ at the collision energy
$\sqrt s= 1$ TeV.

\noindent {\bf Figure 4.} Feynman diagrams for the  production of the doubly
charged higgsino in
electron-electron collisions.

\noindent {\bf Figure 5.} Total cross section for the reaction $e^-e^-\to
\tilde\Delta^{--}\tilde\chi^0$ as a function of the higgsino mass
$m_{\tilde\Delta^{++}}$
for two values of the selectron mass $m_{\tilde l}$ at the collision energy
$\sqrt s= 1$ TeV.

\noindent {\bf Figure 6.} Feynman diagram for the  photoproduction of the
doubly charged higgsino.

\noindent {\bf Figure 7.} Total cross section for the reaction $\gamma e^-\to
\tilde\Delta^{--}\tilde l^+$ as a function of the higgsino mass
$m_{\tilde\Delta^{++}}$
for two values of the selectron mass $m_{\tilde l}$ at the electron-electron
(positron) collision
energy $\sqrt s_{e}= 1$ TeV.

\noindent {\bf Figure 8.} Feynman diagram for the  production of the doubly
charged higgsinos in
photon photon collision.

\noindent {\bf Figure 9.} Total cross section for the reaction $\gamma
\gamma\to
\tilde\Delta^{--}\tilde\Delta^{++}$ as a function of the higgsino mass
$m_{\tilde\Delta^{++}}$
 at the electron-electron  collision
energy $\sqrt s_{ee}= 1$ TeV.

 \begin{table}
\begin{tabular}{|l|c|}\hline
Superfield &
$
{\begin{array}{c}
{\rm Transformation}\: {\rm under} \\
SU(3)_c\times SU(2)_L\times SU(2)_R\times U(1)_{B-L}
\end{array}}$ \\ \hline
{Higgs superfields:}&\\
$ \widehat\phi_u=\left(
{\begin{array}{cc} \widehat\phi_1^0 & \widehat\phi_1^+ \\
\widehat\phi_2^- & \widehat\phi_2^0 \end{array}} \right)_u $ & (1,2,2,0) \\
$\widehat\phi_d=\left(
{\begin{array}{cc} \widehat\phi_1^0 & \widehat\phi_1^+ \\
\widehat\phi_2^- & \widehat\phi_2^0 \end{array}} \right)_d $ & (1,2,2,0) \\
$\widehat\Delta=\left(
{\begin{array}{cc} {\frac 1 {\sqrt{2}}} \widehat\Delta^+ &
\widehat\Delta^{++} \\
\widehat\Delta^0 & - {\frac 1{\sqrt{2}}} \widehat\Delta^+ \end{array}}
\right) $ & (1,1,3,2) \\
$\widehat\delta=\left( {\begin{array}{cc} {\frac 1 {\sqrt{2}}}
\widehat\delta^- & \widehat\delta^{0} \\
\widehat\delta^{--} & - {\frac 1{\sqrt{2}}}
\widehat\delta^- \end{array}} \right) $
& (1,1,3,-2) \\
{superfields containing quarks and leptons:}& \\
$\widehat Q_{Li}=\left( {\begin{array}{c} \widehat u_{Li}
\\ \widehat d_{Li} \end{array}} \right) $ & (3,2,1,1/3) \\
$\widehat Q_{Ri}^c=\left( {\begin{array}{c} \widehat d_{Ri}^c
\\\widehat u_{Ri}^c \end{array}} \right) $ & ($3^*$,1,2,-1/3) \\
$\widehat L_{Li}=\left( {\begin{array}{c} \widehat\nu_{Li}
\\\widehat e_{Li} \end{array}} \right) $ & (1,2,1,-1) \\
$\widehat L_{Ri}^c=\left( {\begin{array}{c} \widehat e_{Ri}^c
\\ \widehat\nu_{Ri}^c \end{array}} \right) $ & (1,1,2,1) \\
{\rm gauge superfields:}& \\
$\widehat G$ & (8,1,1,0)\\
$\widehat W_L$ & (1,3,1,0) \\
$\widehat W_R$ & (1,1,3,0) \\
$\widehat V$ & (1,1,1,0) \\     \hline
\end{tabular}
\bigskip
\caption{The superfields of the supersymmetric left-right model.}
\end{table}

\begin{table}
\begin{tabular}{|l|r|} \hline
chargino $\chi_i^-$  (Dirac fermion, 4-component notation)
& $m_{\chi_i^-}$ [GeV] \\ \hline
$\left( \begin{array}{c} -0.99i\lambda_R^-
-0.07\tilde\phi^-_{2d}+0.09\tilde\delta^- \\
{0.81 \overline{i\lambda_R^+}+0.08\overline{\tilde\phi_{1u}^+}
-0.58\overline{\tilde\Delta^+} } \end{array}\right)$
& 1230  \\
$\left( \begin{array}{c} -i\lambda_L^- +0.1\tilde\phi_{2u}^- \\
{- \overline{i\lambda_L^+}+0.1\overline{\tilde\phi_{1d}^+} }\end{array}\right)$
& 1008  \\
$\left( \begin{array}{c} 0.02i\lambda_R^--0.89\tilde\phi_{2d}^-
-0.45\tilde\delta^- \\
{-0.26 \overline{i\lambda_R^+}-0.84\overline{\tilde\phi_{1u}^+}-
0.48\overline{\tilde\Delta^+ }} \end{array}\right)$
& 212   \\
$\left( \begin{array}{c} -0.1i\lambda_L^- -\tilde\phi_{2u}^- \\
{- 0.1\overline{i\lambda_L^+}-\overline{\tilde\phi_{1d}^+}} \end{array}\right)$
& 192  \\
$\left( \begin{array}{c} 0.11i\lambda_R^-
-0.45\tilde\phi_{2d}^-+0.89\tilde\delta^
- \\
{0.52 \overline{i\lambda_R^+}-0.54\overline{\tilde\phi_{1u}^+}
+0.66\overline{\tilde\Delta^+} } \end{array}\right)$
& 149 \\
\hline
neutralino $\chi_i^0$ (Majorana fermion, 2-component notation)
& $m_{\chi_i^0}$ [GeV]  \\ \hline
$
-0.01i\lambda_R^0+0.73i\lambda_L^0-0.47i\lambda_{B-L}^0+0.04\tilde\phi_{1u}^0-0.01\tilde\phi_{2d}^0
+0.50\tilde\Delta^0+0.07\tilde\delta^0
$ &1479 \\ $
0.87i\lambda_R^0-0.25i\lambda_L^0-0.40i\lambda_{B-L}^0-0.09\tilde\phi_{1u}^0
+0.03\tilde\phi_{2d}^0 +0.01\tilde\Delta^0+0.002\tilde\delta^0$ & 1009 \\
$ 0.48i\lambda_R^0+0.48i\lambda_L^0+0.74i\lambda_{B-L}^0$ & 1000 \\
$
-0.003i\lambda_R^0+0.38i\lambda_L^0-0.24i\lambda_{B-L}^0-0.07\tilde\phi^0_{1u}-0.02\tilde\phi^0_{2d}
-0.83\tilde\Delta^0+0.31\tilde\delta^0 $ &529 \\
$
0.04i\lambda_R^0-0.01i\lambda_L^0-0.02i\lambda_{B-L}^0+0.71\tilde\phi^0_{1u}+0.70\tilde\phi^0_{2d}
-0.05\tilde\Delta^0+0.05\tilde\delta^0 $
&203 \\
$ \tilde\phi^0_{1d}$
&200 \\
$ \tilde\phi^0_{2u}$
&200 \\
$
0.08i\lambda_R^0-0.02i\lambda_L^0-0.03i\lambda_{B-L}^0+0.69\tilde\phi_{1u}^0-0.71\tilde\phi_{2d}^0
-0.06\tilde\Delta^0-0.06\tilde\delta^0
$
&194 \\
$
-0.002i\lambda_R^0+0.18i\lambda_L^0-0.11i\lambda_{B-L}^0-0.03\tilde\phi_{1u}^0+0.08\tilde\phi_{2d}^0
-0.24\tilde\Delta^0-0.94\tilde\delta^0
 $
&51 \\ \hline
\end{tabular}
\bigskip
\caption{Physical charginos and neutralinos for
$m_{W_R}=500\,$GeV (or $v=759\,$GeV), $\mu_1=\mu_2=200$ GeV and soft gaugino
masses of 1 TeV}
\end{table}

\end{document}